\documentstyle[12pt]{article}
\newcommand\bea{\begin{eqnarray}}
\newcommand\eea{\end{eqnarray}}
\begin{document}
\bibliographystyle{unsrt}
\setlength{\baselineskip}{18pt}
\parindent 24pt


\begin{center}{
{\Large {{\bf Purity of states in the theory of open
quantum systems }} }
\vskip 0.5truecm
A. Isar$^*$\\
{\it Department of Theoretical Physics, Institute of
Physics and Nuclear Engineering\\
Bucharest-Magurele, Romania }\\
}
\end{center}

\begin{abstract}

The condition of purity of states for a damped harmonic oscillator is
considered in the framework of Lindblad theory for open quantum systems.
For a special choice of the environment coefficients, the correlated
coherent states are shown to be the only states which remain pure all the time
during the evolution of the considered system.
These states are also the most stable under evolution in the presence of the
environment.
\end{abstract}



* e-mail address: isar@theory.nipne.ro


\section{Introduction}

In the last two decades, more and more interest has arisen about the search
for a consistent description of open quantum systems [1--5]
(for a recent review
see ref. \cite{rev}).
Dissipation
in an open system results from microscopic reversible interactions between
the observable system and the environment.
Because dissipative processes imply irreversibility and,
therefore, a preferred direction in time, it is generally thought that quantum
dynamical semigroups are the basic tools to introduce dissipation in quantum
mechanics. In the Markov approximation
the most general form of the generators of such semigroups was given
by Lindblad \cite{l1}. This formalism has been studied for the case of damped
harmonic oscillators [6, 8--12]
and applied to various physical phenomena,
for
instance, to the damping of collective modes in deep inelastic collisions in
nuclear physics \cite{i1}. A phase space representation for the open quantum
systems within the Lindblad theory was given in \cite{i2,vlas}.

In the present study we are also concerned
with the observable system of a harmonic oscillator which interacts with
an environment. We discuss under what conditions the open system can be
described by a quantum mechanical pure state. In Sec. 2 we present the
generalized uncertainty relations and
the correlated coherent
states, first introduced in \cite{dodkur}, which minimize these relations.
The Lindblad theory for open quantum systems is considered in Sec. 3.
Then in Sec. 4, for the one-dimensional harmonic oscillator as an open system,
we show for a special choice of the diffusion
coefficients that the correlated coherent states, taken as initial states,
remain pure for all time during the evolution.
In some other simple models of the damped harmonic oscillator
in the framework of quantum
statistical theory \cite{zel,hug}, it was shown that the pure
Glauber coherent states remain as those during the evolution
and in all other cases, the oscillator immediately evolves into mixtures.
In this respect
we generalize this result and also our previous result from
\cite{ass} as well as the results of other authors
\cite{halzou}, obtained by using different methods.
In Sec. 5 we introduce the linear entropy and present its role
for the description of the decoherence phenomenon. We claim that the
correlated coherent states are the most stable under evolution
in the presence of the environment and make the connection with the
work done in this field by other authors [20--24].
Finally, concluding remarks are given in Sec. 6.

\section {Generalized uncertainty relations}

In the following we denote by $\sigma_{AA}$ the dispersion of the operator
$\hat A$, i.e. $\sigma_{AA}=<\hat A^2>-<\hat A>^2,$
where $<\hat A>\equiv\sigma_A={\rm Tr}(\hat\rho \hat A), {\rm Tr \hat\rho}=1$
and $\hat\rho$ is the statistical operator (density matrix).
By $\sigma_{AB}=1/2<\hat A\hat B+\hat B\hat A>-<\hat A><\hat B>$ we denote
the correlation of the operators $\hat A$ and $\hat B.$
Schr\" odinger \cite{schr} and Robertson \cite{rob} proved for any
Hermitian operators
$\hat A$ and $\hat B$ and for pure quantum states
the following generalized  uncertainty relation :
\bea \sigma_{AA}\sigma_{BB}-\sigma_{AB}^2\ge {1\over 4}|<[\hat A,\hat B]>|^2.
\label{rob}\eea
For the particular case of the operators of the coordinate $\hat q$
and momentum $\hat p$ the uncertainty relation (\ref{rob}) takes the form
\bea \sigma_{pp}\sigma_{qq}-\sigma_{pq}^2 \ge{\hbar^2\over 4}.\label{genun}\eea
This result was generalized for arbitrary operators (in general non-Hermitian)
and for the most general case of mixed states in \cite{dodkur}.
The inequality (\ref{genun}) can also be represented in the following form
\cite{dodkur}:
\bea \sigma_{pp}\sigma_{qq}\ge{\hbar^2\over 4(1-r^2)},\label{dod}\eea
where
\bea r={\sigma_{pq}\over\sqrt{\sigma_{pp}\sigma_{qq}}} \label{corcoe}\eea
is the correlation coefficient.
The equality in the relation (\ref{genun}) is realized for a special class of
pure states, called correlated
coherent states \cite{dodkur}, which are represented by Gaussian wave packets
in the coordinate representation.
These minimizing states, which generalize
the Glauber coherent states, are eigenstates of an operator of the form
\cite{dodkur}:
\bea \hat a_{r,\eta}={1\over 2\eta}[1-{ir\over (1-r^2)^{1/2}}]\hat q+
i{\eta\over\hbar}\hat p \label{eigv}\eea
with real parameters $r$ and $\eta,$ $|r|<1, \eta=\sqrt{\sigma_{qq}}.$
Their
normalized eigenfunctions, the correlated coherent states, have the form
\cite{dodkur}:
\bea \Psi(x)={1\over (2\pi\eta^2)^{1/4}}\exp\{-{x^2\over 4\eta^2}[1-
{ir\over (1-r^2)^{1/2}}]+{\alpha x\over\eta}-{1\over 2}(\alpha^2+
|\alpha|^2)\}, \label{eigf} \eea
with $\alpha$ a complex number.
If we set $r=0$ and $\eta=(\hbar/2m\omega)^{1/2},$ where $m$ and $\omega$ are
the mass and respectively the frequency of the harmonic oscillator,
the states (\ref{eigf}) become the usual
Glauber coherent states. In Wigner representation, the states (\ref{eigf})
have   the form \cite{dodkur}:
\bea W_{\alpha,r,\eta}(p,q)=
{1\over\pi\hbar}\exp[-{2\eta^2\over\hbar^2}
(p-\sigma_p)^2-{(q-\sigma_q)^2\over 2\eta^2(1-r^2)}+{2r\over \hbar(1-r^2)^
{1/2}}(q-\sigma_q)(p-\sigma_p)]. \label{corwig}\eea
This is the classical normal distribution to give dispersion
\bea \sigma_{qq}=\eta^2,~~ \sigma_{pp}={\hbar^2\over 4\eta^2(1-r^2)},~~
\sigma_{pq}={\hbar r\over 2({1-r^2})^{1/2}} \label{disp} \eea
and the correlation coefficient $r.$
The Gaussian distribution (\ref{corwig}) is the only positive Wigner
distribution for a pure state \cite{huds}. All other Wigner functions
that describe pure states necessarily take on negative values for some
values of $p,q.$

In the case of the relation (\ref{rob}) the equality is generally obtained
only for pure states \cite{dodkur}. For any density matrix in the coordinate
representation (normalized to unity)
the following relation must be fulfilled:
\bea \gamma={\rm Tr}\hat\rho^2\leq 1.
\label{ropur}\eea
The quantity $\gamma$ characterizes the degree of purity of the state.
For pure states $\gamma=1,$ for highly mixed states $\gamma\ll 1$
and for weekly mixed states $1-\gamma\ll 1.$

The Wigner function may be expressed as the Fourier transform of the
off-diagonal matrix elements of the density operator in the coordinate
representation:
\bea W(p,q,t)={1\over \pi\hbar}\int dy<q-y|\hat\rho|q+y>e^{2ipy/\hbar}.
\label{fouri} \eea
Then $<x|\hat\rho|y>$ can be obtained from the inverse Fourier
transform of the Wigner function:
\bea <x|\hat\rho|y>=\int dp \exp({i\over \hbar}p(x-y))W(p,{x+y\over 2}).
\label{fourinv}\eea
In the language of the Wigner function the condition (\ref{ropur})
has the form:
\bea \gamma=2\pi\hbar\int W^2(p,q)dpdq\leq 1.
\label{wigpur}\eea

Let us consider the most general mixed squeezed states described by the
Wigner function of the generic Gaussian form with five real parameters:
\bea   W(p,q)={1\over 2\pi\sqrt{\sigma}}\exp\{-{1\over 2\sigma}[\sigma_{pp}
(q-\sigma_q)^2+\sigma_{qq}(p-\sigma_p)^2-2\sigma_{pq}(q-\sigma_q)(p-\sigma_p)]
\},  \label{gaus}  \eea
where $\sigma$ is the determinant of the dispersion (correlation)
matrix $M$,
\bea\sigma=\det M =\sigma_{pp}\sigma_{qq}-\sigma_{pq}^2\label{det}\eea
and
\bea M=\left(\matrix{\sigma_{pp}&\sigma_{pq}\cr
\sigma_{pq}&\sigma_{qq}\cr}\right). \label{sigma}\eea
The Gaussian Wigner functions of this form correspond to the so-called
quasi-free states on the $C^*$-algebra of the canonical commutation relations,
which is the most natural framework for a unified treatment of quantum and
thermal fluctuations \cite{scut}.
For Gaussian states of the form (\ref{gaus}) the coefficient of purity
$\gamma$ is given by
\bea \gamma={\hbar\over 2\sqrt{\sigma}}. \label{pursig}\eea
Therefore, the dispersion matrix has to
satisfy the Schr\"odinger-Robertson uncertainty relation (\ref{genun})
\bea\sigma\geq {h^2\over 4}.\label{detsig}\eea
This inequality must be fulfilled actually for any states, not only Gaussian.
Any Gaussian pure state minimizes the relation (\ref{detsig}).
Here, $\sigma$ is also the Wigner function area -- a measure of the
phase space area in which the Gaussian density matrix is localized.
For $\sigma>\hbar^2/4$ the function (\ref{gaus}) corresponds to mixed quantum
states, while in the case of the equality $\sigma=\hbar^2/4$ it
takes the form (\ref{corwig}) corresponding
to pure correlated coherent states.

The degree of the purity of a state can also be characterized by other
quantities besides $\gamma.$ The most usual one is the quantum entropy
(we put the Boltzmann's constant $k=1$):
\bea S=-{\rm Tr}(\hat\rho\ln\hat\rho)=-<\ln\hat\rho>.\label{entro}\eea
For quantum pure states the entropy is
identically equal to zero.
It was shown \cite{aur1,aga} that for Gaussian states with the Wigner
functions (\ref{gaus})
the entropy can be expressed through $\sigma$ only:
\bea S=(\nu+1)\ln(\nu+1)-\nu\ln\nu,~~~\nu={1\over\hbar}\sqrt{\sigma}-{1\over 2}.
\label{entro2}\eea

\section{Quantum Markovian master equation for damped
harmonic oscillator}

The simplest dynamics for an open system which describes an irreversible
process is a
semigroup of transformations introducing a preferred direction in time
\cite{d,s,l1}.
In Lindblad's axiomatic formalism of introducing dissipation in quantum
mechanics, the usual von Neumann-Liouville equation ruling the time evolution
of closed quantum systems is replaced by the following Markovian master
equation for the density operator $\hat\rho(t)$ in the Schr\"odinger picture
\cite{l1}:
\bea{d\Phi_{t}(\hat\rho)\over dt}=L(\Phi_{t}(\hat\rho)). \label{lin1}\eea
Here, $\Phi_{t}$ denotes the dynamical semigroup describing the irreversible
time evolution of the open system in the Schr\"odinger representation and $L$
is the infinitesimal generator of the dynamical semigroup $\Phi_t$. Using the
Lindblad theorem \cite{l1} which gives the most general form of a bounded,
completely
dissipative generator $L$, we obtain the explicit form of the most general
quantum mechanical master equation of Markovian type:
\bea{d\hat\rho(t)\over dt}=-{i\over\hbar}[\hat H,\hat\rho(t)]+{1\over 2\hbar}
\sum_{j}([\hat V_{j}\hat\rho(t),\hat V_{j}^\dagger ]+[\hat
V_{j},\hat\rho(t)\hat V_{j}^\dagger ]).\label{lineq}\eea
Here $\hat H$ is the Hamiltonian operator of the system and $\hat V_{j},$
$\hat V_{j}^\dagger $ are bounded operators on the Hilbert space $\cal H$
of the Hamiltonian and they model the environment.
We make the basic assumption that the general
form (\ref{lineq}) of the master equation with a bounded generator
is also valid for
an unbounded generator.
In the case of an exactly solvable model for the damped
harmonic oscillator, we define the possible two operators $\hat V_{1}$ and
$\hat V_{2},$ which are linear in $\hat p$ and $\hat q,$ as follows
\cite{rev,l2,ss}:
\bea \hat V_{j}=a_{j}\hat p+b_{j}\hat q,~j=1,2, \label{oper}\eea
with $a_{j},b_{j}$ complex numbers. The harmonic oscillator
Hamiltonian $\hat H$ is chosen of the general form
\bea   \hat H=\hat H_{0}+{\mu \over 2}(\hat q\hat p+\hat p\hat q),
~~~\hat H_{0}={1\over 2m}
\hat p^2+{m\omega^2\over 2}\hat q^2.  \label{ham}   \eea
With these choices and with the notations
\bea D_{qq}={\hbar\over 2}\sum_{j=1,2}{\vert a_{j}\vert}^2,
  D_{pp}={\hbar\over 2}\sum_{j=1,2}{\vert b_{j}\vert}^2,
D_{pq}=D_{qp}=-{\hbar\over 2}{\rm Re}\sum_{j=1,2}a_{j}^*b_{j},
\lambda=-{\rm Im}\sum_{j=1,2}a_{j}^*b_{j},\label{coef}\eea
where $a_j^*$ and $b_j^*$ denote the complex conjugate of $a_j,$
and $b_j,$ respectively,
the master equation (\ref{lineq}) takes the following form \cite{rev,ss}:
\bea   {d\hat\rho \over dt}=-{i\over \hbar}[\hat H_{0},\hat\rho]-
{i\over 2\hbar}(\lambda +\mu)
[\hat q,\hat\rho \hat p+\hat p\hat\rho]+{i\over 2\hbar}(\lambda -\mu)[\hat p,
\hat\rho \hat q+\hat q\hat\rho]  \nonumber\\
  -{D_{pp}\over {\hbar}^2}[\hat q,[\hat q,\hat\rho]]-{D_{qq}\over {\hbar}^2}
[\hat p,[\hat p,\hat\rho]]+{D_{pq}\over {\hbar}^2}([\hat q,[\hat p,\hat\rho]]+
[\hat p,[\hat q,\hat\rho]]). ~~~~\label{mast}   \eea
Here the quantum mechanical diffusion coefficients $D_{pp},D_{qq},$ $D_{pq}$
and the friction constant $\lambda$ satisfy the following fundamental
constraints \cite{rev,ss}: $  D_{pp}>0, D_{qq}>0$ and
\bea D_{pp}D_{qq}-D_{pq}^2\ge
{{\lambda}^2{\hbar}^2\over 4}.  \label{ineq}  \eea
The relation (\ref{ineq}) is a necessary condition for the
generalized uncertainty inequality (\ref{genun}) to be fulfilled.

The semigroup method is valid for the weak-coupling regime, with the damping
$\lambda$ typically obeying the inequality $\lambda\ll\omega_0,$ where
$\omega_0$ is the lowest frequency typical of reversible motions.

The necessary and sufficient condition for $L$ to be translationally
invariant is $\mu=\lambda$ \cite{rev,l2,ss}.
Translation invariance means that $[p,L(\rho)]=L([p,\rho]).$
In the following general values
for $\lambda$ and $\mu$ will be considered.

By using the fact that
the linear positive mapping
defined by $\hat A\to {\rm Tr}(\hat\rho\hat A)$
is completely positive,  in \cite{rev,ss} the
following inequality was obtained:
\bea D_{pp}\sigma_{qq}(t)+D_{qq}\sigma_{pp}(t)-2D_{pq}\sigma_{pq}(t)\ge
{\hbar^2\lambda\over 2}\label{has3}.\eea
We have found in \cite{ass} that this inequality, which must be valid for all
values of $t$ is equivalent with the generalized
uncertainty inequality (\ref{genun}) at any time $t,$
\bea\sigma_{qq}(t)\sigma_{pp}(t)-\sigma_{pq}^2(t)\ge{\hbar^2
\over 4},\label{genun1}\eea
if the initial values $\sigma_{qq}(0),\sigma_
{pp}(0)$ and $\sigma_{pq}(0)$ for $t=0$ satisfy this inequality.
If the initial state is the ground state of the harmonic oscillator, then
\bea\sigma_{qq}(0)={\hbar\over 2m\omega},~\sigma_{pp}(0)={m\hbar\omega\over 2},
~\sigma_{pq}(0)=0. \label{grovar1}\eea

By using the complete positivity property of the dynamical semigroup $\Phi_t,$
it was shown in \cite{ss} that the relation
\bea{\rm Tr}(\Phi_t(\hat\rho)\sum_j\hat V_j^\dagger\hat V_j)=\sum_j{\rm Tr}
(\Phi_t(\hat\rho)\hat V_j^\dagger){\rm Tr}(\Phi_t(\hat\rho)\hat V_j)
\label{has1}\eea
represents the necessary and sufficient condition for $\hat\rho(t)=
\Phi_t(\hat\rho)$ to be a pure state for all times $t\ge 0.$
This equality is a generalization of the pure state condition [30--32]
to all Markovian master equations (\ref{lineq}).
If $\hat\rho^2(t)=\hat\rho(t)$ for all $t\ge 0,$ there exists a
wave function
$\psi\in{\cal H}$ which satisfies the nonlinear Schr\"odinger-type equation
\bea i\hbar{\partial \psi(t)\over \partial t}=\hat H'\psi(t), \label{scheq}
\eea
with the non-Hermitian Hamiltonian
\bea \hat H'=\hat H+i\sum_j<\psi(t),\hat V_j^\dagger\psi(t)>\hat V_j-{i\over 2}
<\psi(t),\sum_j\hat V_j^\dagger\hat V_j\psi(t)>-{i\over 2}\sum_j
\hat V_j^\dagger\hat V_j. \label{ham1} \eea
For environment operators $\hat V_j$ of the form (\ref{oper}),
the pure state condition (\ref{has1}) takes the following form
\cite{ss}, corresponding to equality in the relation (\ref{has3}):
\bea D_{pp}\sigma_{qq}(t)+D_{qq}\sigma_{pp}(t)-2D_{pq}\sigma_{pq}(t)=
{\hbar^2\lambda\over 2}\label{has2}\eea
and the Hamiltonian (\ref{ham1}) becomes
\bea \hat H'=\hat H+\lambda(\sigma_p(t)\hat q-
\sigma_q(t)\hat p)+{i\over\hbar}[\lambda\hbar^2-D_{qq}((\hat p-
\sigma_p(t))^2+
\sigma_{pp}(t))
-D_{pp}((\hat q-
\sigma_q(t))^2\nonumber \\
+\sigma_{qq}(t))
+D_{pq}((\hat p-
\sigma_p(t))(\hat q-
\sigma_q(t))+(\hat q-
\sigma_q(t))(\hat p-
\sigma_p(t))+2
\sigma_{pq}(t))]. ~~~~~~~\label{ham2}\eea

From the master equation (\ref{mast}) we obtain the following equations of
motion for the expectation values and variances of the coordinate and momentum:
\bea{d\sigma_{q}(t)\over dt}=-(\lambda-\mu)\sigma_{q}(t)+{1\over m}\sigma_{p}
(t),\eea
\bea{d\sigma_{p}(t)\over dt}=-m\omega^2\sigma_{q}(t)-(\lambda+\mu)\sigma_{p}
(t) \label{eqmo1}\eea
and
\bea{d\sigma_{qq}(t)\over dt}=-2(\lambda-\mu)\sigma_{qq}(t)+{2\over m}
\sigma_{pq}(t)+2D_{qq},\eea
\bea{d\sigma_{pp}\over dt}=-2(\lambda+\mu)\sigma_{pp}(t)-2m\omega^2
\sigma_{pq}(t)
+2D_{pp}, \label{eqmo2}\eea
\bea{d\sigma_{pq}(t)\over dt}=-m\omega^2\sigma_{qq}(t)+{1\over m}\sigma_{pp}(t)
-2\lambda\sigma_{pq}(t)+2D_{pq}.\eea
In the underdamped case $(\omega>\mu)$ considered in this paper, with
the notation $\Omega^2=\omega^2-\mu^2$, we obtain \cite{rev,ss}:
\bea\sigma_q(t)=e^{-\lambda t}((\cos\Omega t+{\mu\over\Omega}\sin\Omega t)
\sigma_q(0)+{1\over m\Omega}\sin\Omega t\sigma_p(0)), \label{sol}\eea
\bea\sigma_p(t)=e^{-\lambda t}(-{m\omega^2\over\Omega}\sin\Omega t\sigma_q(0)+
(\cos\Omega t-{\mu\over\Omega}\sin\Omega t)\sigma_p(0)) \label{sol1}\eea
and $\sigma_q(\infty)=\sigma_p(\infty)=0.$
It is convenient to consider the vectors
\bea X(t)=\left(\matrix{m\omega\sigma_{qq}(t)\cr
\sigma_{pp}(t)/m\omega\cr
\sigma_{pq}(t)\cr}\right)\eea
and
\bea D=\left(\matrix{2m\omega D_{qq}\cr
2D_{pp}/m\omega\cr
2D_{pq}\cr}\right).\eea
With these notations the solutions for the variances can be written in the
form \cite{rev,ss}:
\bea X(t)=(Te^{Kt}T)(X(0)-X(\infty))+X(\infty),\label{sol2}\eea
where the matrices $T$ and $K$ are given by
\bea T={1\over 2i\Omega}\left(\matrix{\mu+i\Omega&\mu-i\Omega&2\omega\cr
\mu-i\Omega&\mu+i\Omega&2\omega\cr
-\omega&-\omega&-2\mu\cr}\right),\eea
\bea K=\left(\matrix{-2(\lambda-i\Omega)&0&0\cr
0&-2(\lambda+i\Omega)&0\cr
0&0&-2\lambda\cr}\right)\label{matr}\eea
and
\bea X(\infty)=-(TK^{-1}T)D.\label{xinf}\eea
The formula (\ref{xinf}) is remarkable because it gives a very simple
connection
between the asymptotic values $(t \to \infty)$ of $\sigma_{qq}(t),
\sigma_{pp}(t), \sigma_{pq}(t)$
and the diffusion coefficients $D_{qq},~D_{pp},~ $ $D_{pq}:$
\bea\sigma_{qq}(\infty)={1\over 2(m\omega)^2\lambda(\lambda^2+\omega^2-\mu^2)}
((m\omega)^2(2\lambda(\lambda+\mu)+\omega^2)D_{qq}\nonumber\\
+\omega^2D_{pp}+2m\omega^2(\lambda+\mu)D_{pq}), ~~~~~~~~~~~~~~~~~~~~~\eea
\bea\sigma_{pp}(\infty)={1\over 2\lambda(\lambda^2+\omega^2-\mu^2)}((m\omega)^2
\omega^2D_{qq}+(2\lambda(\lambda-\mu)+\omega^2)D_{pp}-2m\omega^2(\lambda-
\mu)D_{pq}),\label{varinf}\eea
\bea\sigma_{pq}(\infty)={1\over 2m\lambda(\lambda^2+\omega^2-\mu^2)}(-(\lambda+
\mu)(m\omega)^2D_{qq}+(\lambda-\mu)D_{pp}+2m(\lambda^2-\mu^2)D_{pq}).\eea
These relations show that the asymptotic values $\sigma_{qq}(\infty),
\sigma_{pp}(\infty),\sigma_{pq}(\infty)$ do not depend on the initial values
$\sigma_{qq}(0),\sigma_{pp}(0),\sigma_{pq}(0)$.
In the considered underdamped case
we have
\bea Te^{Kt}T=-{e^{-2\lambda t}\over 2\Omega^2}
\left(\matrix{b_{11}&b_{12}&b_{13}\cr
b_{21}&b_{22}&b_{23}\cr
b_{31}&b_{32}&b_{33}\cr}\right),\eea
where $b_{ij},$ $i,j=1,2,3$ are time-dependent oscillating functions
given by
(3.78) in \cite{ss}.

\section{Purity of states}

We will be interested to find the Gaussian states which remain pure during the
evolution of the system for all times $t.$ We start by considering the
pure state condition (\ref{has2}) and the generalized uncertainty relation
(\ref{genun}) which transforms into the following minimum uncertainty
equality for pure states:
\bea \sigma_{pp}(t)\sigma_{qq}(t)-\sigma_{pq}^2(t) ={\hbar^2\over 4}.
\label{pur}\eea
By eliminating $\sigma_{pp}$ between the equalities (\ref{has2}) and
(\ref{pur}), like in \cite{deval}, we obtain:
\bea(\sigma_{qq}(t)-{D_{pq}\sigma_{pq}(t)+{1\over 4}\hbar^2\lambda\over
D_{pp}})^2+{D_{pp}D_{qq}-D_{pq}^2\over D_{pp}^2}[(\sigma_{pq}(t)-{{1\over 4}
\hbar^2\lambda D_{pq}\over D_{pp}D_{qq}-D_{pq}^2})^2 \nonumber\\
+{1\over 4}\hbar^2{D_{pp}D_{qq}-D_{pq}^2-{1\over 4}\hbar^2\lambda^2\over
(D_{pp}D_{qq}-D_{pq}^2)^2}D_{pp}D_{qq}]=0.~~~~~~~~~~~~~~~~~~ \label{dek}\eea
Since the opening coefficients satisfy the inequality (\ref{ineq}),
from the relation (\ref{dek}) we obtain the following relations
which have to be fulfilled at any moment of time:
\bea D_{pp}D_{qq}-D_{pq}^2={\hbar^2\lambda^2\over 4}, \label{coe1}\eea
\bea D_{pp}\sigma_{qq}(t)-D_{pq}\sigma_{pq}(t)-{\hbar^2\lambda\over 4}=0,
\label{coe2}\eea
\bea \sigma_{pq}(t)(D_{pp}D_{qq}-D_{pq}^2)-{\hbar^2\lambda\over 4}D_{pq}=0.
\label{coe3}\eea
From the relations (\ref{pur}) and (\ref{coe1}) -- (\ref{coe3}) it
follows that the pure states remain pure for all times only if
their variances are constant in time and have the form:
\bea \sigma_{pp}(t)={D_{pp}\over\lambda},~~
\sigma_{qq}(t)={D_{qq}\over\lambda},~~
\sigma_{pq}(t)={D_{pq}\over\lambda}. \label{coe4}\eea
If these relations are fulfilled, then the inequalities
(\ref{has3}) and (\ref{genun1}) are both equivalent to (\ref{ineq}),
including also the corresponding equalities (\ref{has2}), (\ref{pur}) and
(\ref{coe1}).
From Eq. (\ref{sol2}) it follows that the variances remain constant
and do not depend on time only if $X(0)=X(\infty),$ which means
$\sigma_{pp}(0)=\sigma_{pp}(\infty),$
$\sigma_{qq}(0)=\sigma_{qq}(\infty),$$\sigma_{pq}(0)=\sigma_{pq}(\infty).$
Using the asymptotic values (48) -- (50) of the variances and the
relations (\ref{coe4}), we obtain
the following expressions of the diffusion coefficients which
assure that the initial pure states remain pure for any $t:$
\bea D_{qq}={\hbar\lambda\over 2m\Omega},~~
D_{pp}={\hbar\lambda m\omega^2 \over 2\Omega},~~
D_{pq}=-{\hbar\lambda\mu\over 2\Omega}. \label{coepur}\eea
Formulas (\ref{coepur}) are generalized Einstein
relations and represent examples of quantum
fluctuation-dissipation relations, connecting the diffusion with both
Planck's constant and damping constant.
With the coefficients (\ref{coepur}), the variances (48) -- (50)
become:
\bea \sigma_{qq}={\hbar\over 2m\Omega},~~
\sigma_{pp}={\hbar m\omega^2 \over 2\Omega},~~
\sigma_{pq}=-{\hbar\mu\over 2\Omega}. \label{varpur}\eea
Therefore, the quantity $\sigma$ (see Eq. (\ref{det})) is equal
to its minimum possible
value $\hbar^2/4,$ according to the generalized uncertainty relation
(\ref{detsig}). Then the corresponding state described by a Gaussian Wigner
function is a pure quantum state, namely a correlated coherent state
\cite{dodkur} with the correlation coefficient (\ref{corcoe}) $r=-\mu/\omega.$
Given $\sigma_{pp},$ $\sigma_{qq}$ and $\sigma_{pq},$ there exists one and only
one such a state minimizing the uncertainty $\sigma$ \cite{sud}.
A particular case of our result (corresponding to $\lambda=\mu$ and $D_{pq}=0)$
was obtained by
Halliwell and Zoupas by using the quantum state diffusion
method \cite{halzou}.
We remark that the minimization of the quantity (\ref{det}) is equivalent,
by virtue of the relation (\ref{entro2}), to the minimization of the
entropy $S$.
As we have mentioned before, we have considered general
coefficients $\mu$ and
$\lambda$ and in this respect our expressions for the diffusion
coefficients and variances generalize also the ones
obtained by Dekker and
Valsakumar \cite{deval} and Dodonov and Man'ko \cite{dodman},
who used models where $\mu=\lambda$ was chosen.
If $\mu=0,$ we get from (\ref{coepur}) $D_{pq}=0.$ This case, which
was considered in \cite{ass}, where we have obtained a density operator
describing a pure state
for any $t,$ is also a particular case of the present results which give the
most general Gaussian pure state which remains pure for any $t.$
For $\mu=0$, the expressions (\ref{varpur}) become
\bea \sigma_{qq}={\hbar\over 2m\omega},~~
\sigma_{pp}={\hbar m\omega\over 2},~~
\sigma_{pq}=0,   \label{grovar2}\eea
which are the values of variances (\ref{grovar1}) for the ground state of
the harmonic oscillator
and the correlation coefficient (\ref{corcoe}) takes the value
$r=0,$ corresponding
to the case of usual coherent states.

The Lindblad equation or its equivalent Fokker-Planck equation for the Wigner
function with the diffusion coefficients (\ref{coepur}) can
be used only in the underdamped case, when $\omega>\mu.$
Indeed, for the coefficients (\ref{coepur}) the fundamental constraint
(\ref{ineq}) implies that $m^2(\omega^2-\mu^2)D_{qq}^2\ge\hbar^2\lambda^2/4,$
which is satisfied only if $\omega>\mu$.
It can be shown that
there exist diffusion coefficients which satisfy the condition
(\ref{coe1}) and make sense for $\omega<\mu,$ but in this overdamped case
we have always $\sigma>\hbar^2/4$ and the state of the oscillator cannot
be pure for any diffusion coefficients \cite{dodman}.

The fluctuation energy of the open harmonic oscillator is
\bea E(t)={1\over 2m}\sigma_{pp}(t)+{1\over 2}m\omega^2\sigma_{qq}(t)+
\mu\sigma_{pq}(t). \label{ener}\eea
If the state remains pure in time, then the variances are given by
(\ref{coe4}) and the fluctuation energy is also constant
in time and is given by
\bea E={1\over\lambda}({1\over 2m}D_{pp}+{1\over 2}m\omega^2D_{qq}+\mu D_{pq}).
\label{conen}\eea
Minimizing this expression with the condition (\ref{coe1}), we obtain
just the diffusion coefficients  (\ref{coepur}) and $E_{min}=\hbar\Omega/2.$
Therefore, the conservation of the purity of state implies
that the fluctuation energy of the system has all the time the minimum
possible value $E_{min}.$ Moreover, we can consider that the case when the
diffusion coefficients satisfy the equality (\ref{coe1}) corresponds to a
zero temperature of the environment (bath). Then the influence on the
oscillator is minimal.

If we choose the coefficients of the form (\ref{coepur}), then
the equation for
the density operator can be represented in the form (\ref{lineq}) with only
one operator $\hat V,$ which up to a phase factor can be written in the form:
\bea \hat V=\sqrt{{2\over\hbar D_{qq}}}[({\lambda\hbar\over 2}-iD_{pq})\hat q+
iD_{qq}\hat p)], \label{oneop}\eea
with $[\hat V,\hat V^\dagger]=2\hbar\lambda.$

The correlated coherent states
(\ref{eigf}) with nonvanishing momentum average, can also be
written in the form:
\bea \Psi(x)=({1\over 2\pi\sigma_{qq}})^{1\over 4}\exp[-{1\over 4\sigma_{qq}}
(1-{2i\over\hbar}\sigma_{pq})(x-\sigma_q)^2+{i\over \hbar}\sigma_px]
\label{ccs}\eea
and the most general form of Gaussian density matrices compatible with the
generalized uncertainty relation (\ref{genun}) is the following:
\bea <x|\hat\rho|y>=({1\over 2\pi\sigma_{qq}})^{1\over 2}
\exp[-{1\over 2\sigma_{qq}}({x+y\over 2}-\sigma_q)^2 ~~~~~~~~~~\nonumber \\
-{1\over 2\hbar^2}
(\sigma_{pp}-{\sigma_{pq}^2\over\sigma_{qq}})(x-y)^2
+{i\sigma_{pq}\over\hbar\sigma_{qq}}({x+y\over 2}-\sigma_q)(x-y)+
{i\over \hbar}\sigma_p(x-y)].\label{ccd}\eea
For these matrices we can verify that
${\rm Tr}\rho^2=\hbar/2\sqrt{\sigma}$
and they correspond to the correlated coherent states (\ref{ccs})
if $\sigma_{pp}, \sigma_{qq}$ and $\sigma_{pq}$ in (\ref{ccd}) satisfy the
equality in (\ref{genun}).

In \cite{a,i2,vlas} we have transformed the master
equation for the density operator into the following Fokker-Planck equation
satisfied by the Wigner
distribution
function $W(q,p,t):$
\bea   {\partial W\over\partial t}=
-{p\over m}{\partial W\over\partial q}
+m\omega^2 q{\partial W\over\partial p}
+(\lambda-\mu){\partial\over\partial q}(qW)
+(\lambda+\mu){\partial\over\partial p}(pW) \nonumber \\
+D_{qq}{\partial^2 W\over\partial q^2}
+D_{pp}{\partial^2 W\over\partial p^2}
+D_{pq}{\partial^2 W\over\partial p\partial q}.~~~~~~~~~~~~~~~~~~~
\label{wigeq}\eea
For an initial Gaussian Wigner function
the solution of Eq. (\ref{wigeq}) is
\bea   W(p,q,t)={1\over 2\pi\sqrt{\sigma}}
~~~~~~~~~~~~~~~~~~~~~~~~~~~~~~~~~ \nonumber \\
\times\exp\{-{1\over 2\sigma}[\sigma_{pp}(q-\sigma_q(t))^2+
\sigma_{qq}(p-\sigma_p(t))^2-2\sigma_{pq}(q-\sigma_q(t))(p-\sigma_p(t))]\}.
\label{wig} \eea
We see that the initial Wigner function remains Gaussian and therefore the
property of positivity is preserved in time.
Consider the harmonic oscillator initially in a correlated coherent state
of the form (\ref{ccs}), with the corresponding Wigner
function (\ref{corwig}).
For our environment described by the diffusion coefficients (\ref{coepur}),
the solution for the Wigner function at time $t$ is given by (\ref{wig}),
where $\sigma_q(t)$ and $\sigma_p(t)$ are given respectively by
(\ref{sol}) and (\ref{sol1}) and the variances by (\ref{varpur}).
Using either (\ref{fourinv}) and (\ref{wig}) or
(\ref{ccd}), we get for the density
matrix the following time evolution:
\bea <x|\hat\rho(t)|y>= ({m\Omega\over\pi \hbar})^{1\over 2}
\exp[-{m\Omega\over \hbar}({x+y\over 2}-\sigma_q(t))^2 ~~~~~~~~~~\nonumber \\
-{m\Omega\over 4\hbar}(x-y)^2
-{im\mu\over \hbar}({x+y\over 2}-\sigma_q(t))(x-y)+{i\over\hbar}\sigma_p(t)
(x-y)]. \label{coord}\eea
In the long time limit $\sigma_q(t)=0,$ $\sigma_p(t)=0$ and we have
\bea <x|\hat\rho(\infty)|y>= ({m\Omega\over\pi \hbar})^{1\over 2}
\exp\{-{m\over 2\hbar}[\Omega(x^2+y^2)+i\mu(x^2-y^2)]\}. \label{coorinf}\eea
The corresponding Wigner function has the form
\bea   W_{\infty}(p,q)={1\over \pi\hbar}
\exp[-{1\over \hbar\Omega}({p^2\over m}+m\omega^2q^2+2\mu qp)].
\label{wiginf}  \eea

We see that the time evolution of the initial correlated coherent state
of the damped harmonic oscillator is given by a Gaussian density matrix with
variances constant in time. According to known general results
\cite{AndH,AnH}, the initial
Gaussian density matrix remains Gaussian centered around the classical path.
So, the correlated coherent state remains a correlated coherent
state and $\sigma_q(t)$ and $\sigma_p(t)$ give the average time dependent
location of the system along its trajectory in phase space.

\section{Entropy and decoherence}

Besides the von Neumann entropy $S$ (\ref{entro}), there is another quantity
which can measure the degree of mixing or purity of quantum states. It
is the linear entropy $S_l$ defined as
\bea S_l={\rm Tr}(\hat\rho-\hat\rho^2)=1-{\rm Tr}\hat\rho^2.\label
{entro3}\eea
For pure states $S_l=0$ and for a statistical mixture $S_l>0.$

As it is well-known, the increasing of the linear entropy $S_l$
(as well as of the von Neumann entropy $S$) due to the interaction with the
environment is associated with the
decoherence phenomenon (loss of quantum coherence), given by the diffusion
process \cite{paz1,paz2}.
Dissipation increases the entropy and the pure states are converted into
mixed states.
The rate of entropy production is given by
\bea {\dot S}_l(\hat\rho)=-2{\rm Tr}(\hat\rho\dot{\hat \rho})=-2{\rm Tr}
(\hat\rho L(\hat\rho)), \label{entpro1}\eea
where $L$ is the evolution operator.
According to Zurek's theory, the maximally predictive states are the pure
states which minimize the entropy production in time.
These states remain least affected by the openness of the system
and form a "preferred set of
states" in the Hilbert space of the system, known as the
"pointer basis". Their evolution is predictible with the principle of least
possible entropy production.

Using (\ref{mast}), in our model the rate of entropy
production (\ref{entpro1}) is given by:

\bea \dot S_l(t)={4\over\hbar^2}
[D_{pp}{\rm Tr}(\hat\rho^2\hat q^2-\hat\rho\hat q\hat\rho\hat q)
~~~~~~~~~~~~~~~~~~~~\nonumber \\
 +D_{qq}{\rm Tr}(\hat\rho^2\hat p^2-\hat\rho\hat p\hat\rho\hat p) -
D_{pq}{\rm Tr}(\hat\rho^2(\hat q\hat p+\hat p\hat q)-
2\hat\rho\hat q\hat\rho\hat p) -
{\hbar^2\lambda\over 2}{\rm Tr}(\hat\rho^2)]. \label{entpro3} \eea
When the state remains approximately pure $(\hat\rho^2\approx\hat\rho),$
we obtain:
\bea \dot S_l(t)={4\over\hbar^2}(D_{pp}\sigma_{qq}(t)+D_{qq}\sigma_{pp}(t)-
2D_{pq}\sigma_{pq}(t)-{\hbar^2\lambda\over 2})\ge 0,\label{entpro4}\eea
according to (\ref{has3}).
We see that $\dot S_l(t)=0$ when the condition (\ref{has2}) of
purity
for any time $t$ is fulfilled. The entropy production $S_l$ is
also equal to 0 at $t=0$ if the initial state is a pure state. But we have shown
before that the only initial states which remain pure for any $t$ are the
correlated coherent states. Therefore, we can state that in the Lindblad
theory for the open quantum harmonic oscillator the correlated
coherent states, which are generalized coherent states, are the maximally
predictive states. Our result generalizes
the previous results which assert that for
many models of quantum Brownian motion in the high
temperature limit the usual coherent states correspond to minimal entropy
production and, therefore, they are the maximally predictive states.
In our model the coherent states can be obtained
as a particular case of the correlated coherent states by taking
$\mu=0,$ so that the correlation coefficient $r=0$ (see Eq. (\ref{corcoe})).

Paz, Habib and Zurek \cite{paz1,paz2}
considered the harmonic oscillator undergoing
quantum Brownian motion in the Caldeira-Leggett model
and concluded that the
minimizing states which are the initial states generating the least amount
of von Neumann or linear entropy and, therefore, the most predictible or
stable ones under evolution in the presence
of an environment are the ordinary coherent states.
Using an information-theoretic measure of uncertainty for quantum systems,
Anderson and Halliwell showed in \cite{AndH} that the minimizing states
are more general Gaussian states. Anastopoulos and Halliwell
\cite{AnH} offered an alternative characterization of these states by noting
that these states
minimize the generalized uncertainty relation.
According to this assertion, we can say that in our model the correlated
coherent states are the most stable states which minimize the
generalized uncertainty relation (\ref{genun}). Our result
confirms the one of
\cite{AnH},
but the model used in \cite{AnH} is different, namely
the open quantum system consists of a particle
moving in a harmonic oscillator potential and is linearly coupled to an
environment consisting of a bath of harmonic oscillators in a thermal
state.
At the same time
we remind that the Caldeira-Leggett model considered in \cite{paz1,paz2}
violates the positivity of the density operator at short time scales
\cite{Amb,dio}, whereas
in the Lindblad model considered here the property of positivity is
automatically fulfilled.

The rate of predictibility loss, measured by the rate of linear entropy
increase, for a damped harmonic oscillator is also calculated in the framework
of Lindblad theory in Ref. \cite{par}. The initial states which minimize
the predictability loss are identified as quasi-free states with a symmetry
dictated by the environment diffusion coefficients. For an isotropic
diffusion in phase space, the coherent states or mixtures of coherent states
are selected as the most stable ones.

In order to generalize the results of Zurek and collaborators, the entropy
production
was considered by Gallis \cite{gal} within the Lindblad theory of open quantum
systems, treating environment effects perturbatively. Gallis considered the
particular case with $D_{pq}=0$ and found out that the squeezed states
emerge as the most stable states for intermediate times compared to the
dynamical time scales. The amount of squeezing decreases with time, so that
the coherent states are most stable for large time scales.
For $D_{pq}\not=0$ we have generalized the result of Gallis and established
that
the correlated coherent states are the most stable under the evolution in the
presence of an environment.

\section{Concluding remarks}

Recently there is a revival of interest in quantum Brownian motion as a
paradigm of quantum open systems. The possibility
of preparing systems in macroscopic quantum states led to the problems of
dissipation in tunneling and of loss of quantum coherence (decoherence). These
problems are intimately related to the issue of quantum to classical
transition and
all of them point the necessity of a better understanding of open quantum
systems.
The Lindblad theory provides a selfconsistent
treatment of damping as a general extension of quantum mechanics to open
systems and gives the possibility to extend the model of quantum Brownian
motion. In the present paper we have studied the one-dimensional harmonic
oscillator with dissipation within the framework of this theory.
We have shown that the only states which stay pure during the evolution
in time of the system are the correlated coherent states under the condition
of a special
choice of the environment coefficients. These states are also connected with
the decoherence phenomenon and they are the most stable under the evolution
in the presence of the environment.
In a future work in the framework of the Lindblad theory we plan to discuss
the connection
between uncertainty, decoherence and correlations of open quantum systems
with their environment in more details.


\end{document}